\documentclass[aps,prl,reprint,showpacs,showkeys,superscriptaddress,preprintnumbers]{revtex4-1}
\usepackage{amsmath,amssymb,bm,mathrsfs}
\usepackage{srcltx,hyperref}
\usepackage{dsfont}
\usepackage{epsfig}
\usepackage{slashed}
\usepackage{bbold}
\usepackage{psfrag}
\usepackage{color}
\PassOptionsToPackage{caption=false}{subfig}
\usepackage{subfig}
\usepackage{multirow}
\usepackage{booktabs}
\usepackage{hyperref}

\begin{document}

\title{Forbidden Dark Matter} 

\author{Raffaele Tito D'Agnolo}
\email{dagnolo@ias.edu}
\affiliation{Institute for Advanced Study, Princeton, NJ 08540, USA}

\author{Joshua T. Ruderman}
\email{ruderman@nyu.edu}
\affiliation{
Center for Cosmology and Particle Physics, \\
Department of Physics, New York University, New York, NY 10003, USA.
}

\begin{abstract}
Dark Matter (DM) may be a thermal relic that annihilates into heavier states in the early Universe.   This Forbidden DM framework accommodates a wide range of DM masses from keV to weak scales.  An exponential hierarchy between the DM mass and the weak scale follows from the exponential suppression of the thermally averaged cross section.  Stringent constraints from the cosmic microwave background are evaded because annihilations turn off at late times.  We provide an example where DM annihilates into dark photons, which is testable through large DM self-interactions and direct detection.
\end{abstract}

\maketitle

\noindent {\bf Introduction:} Dark Matter (DM) accounts for over 80\% of the matter of our Universe, but its particle origin remains a mystery.   
If it is a thermal relic, its energy density today depends on its annihilation rate, $\left< \sigma v \right>$, in the early Universe~\cite{Kolb:1990vq},
\begin{equation} \label{eq:Omega}
\Omega_{DM} h^2 \sim 0.1 \frac{(20~\mathrm{TeV})^{-2}}{\left< \sigma v \right>}.
\end{equation}
The observed DM abundance, $\Omega_{DM} h^2 \sim 0.1$, is reproduced if DM annihilates with a weak scale cross section.   

Typically, $\left< \sigma v \right> \sim \alpha_d^2 / m_{\psi}^2$, where $m_{\psi}$ is the DM mass and $\alpha_d$ characterizes its interaction strength.  For sizable $\alpha_d$, the DM mass should be near the weak scale, $m_{\psi} \sim 0.1 - 10$~TeV\@.   Direct detection experiments have been making rapid progress searching for weak scale DM scattering against nuclei, and significant reach is projected over the next decade (see Ref.~\cite{Cushman:2013zza} for a review). 
However, pending a possible discovery, it is important to cast a wide net by exploring a diverse set of models with varying DM masses and signatures.

There are several classes of models where the  DM mass may be much lighter than the weak scale.  One possibility is to take $\alpha_d \ll 1$, keeping $\alpha_d^2 / m_{\psi}^2$ fixed~\cite{Feng:2008ya}.  A second possibility is that the DM energy density is determined by an asymmetry, similarly to baryons, such that the DM mass varies inversely with the size of the asymmetry~\cite{Nussinov:1985xr,Kaplan:2009ag}.  A third possibility is that the DM relic density is determined by $3 \rightarrow 2$ annihilations instead of $2 \rightarrow 2$ annihilations, requiring light DM to achieve the necessary cross section~\cite{Carlson:1992fn, Hochberg:2014dra}.

\begin{figure*}[tb]
\centering
\begin{tabular}{@{}ll@{}}
\raisebox{.15\height}{\includegraphics[width=0.45\linewidth]{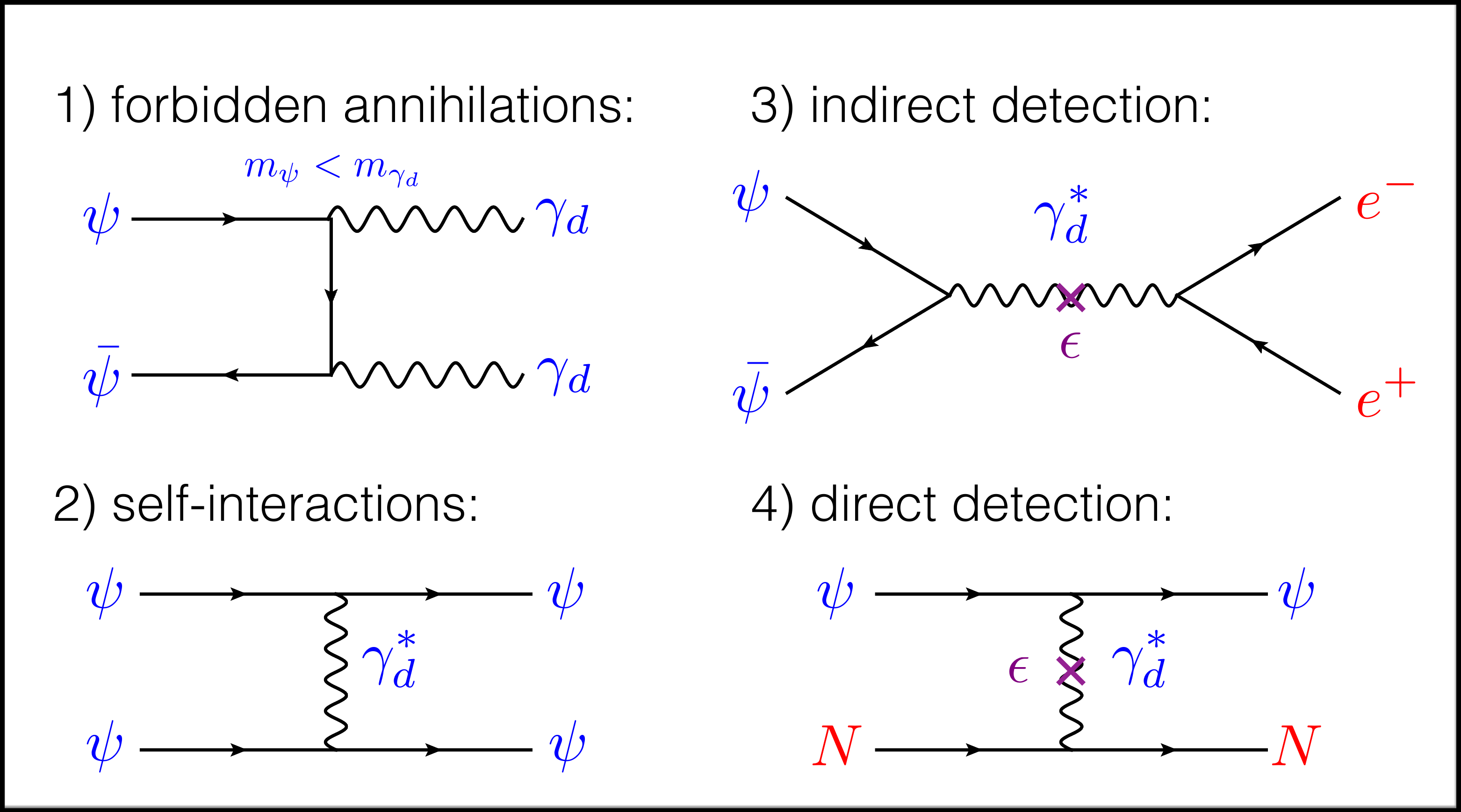}} &
\hspace{0.75cm}\includegraphics[width=0.45\linewidth]{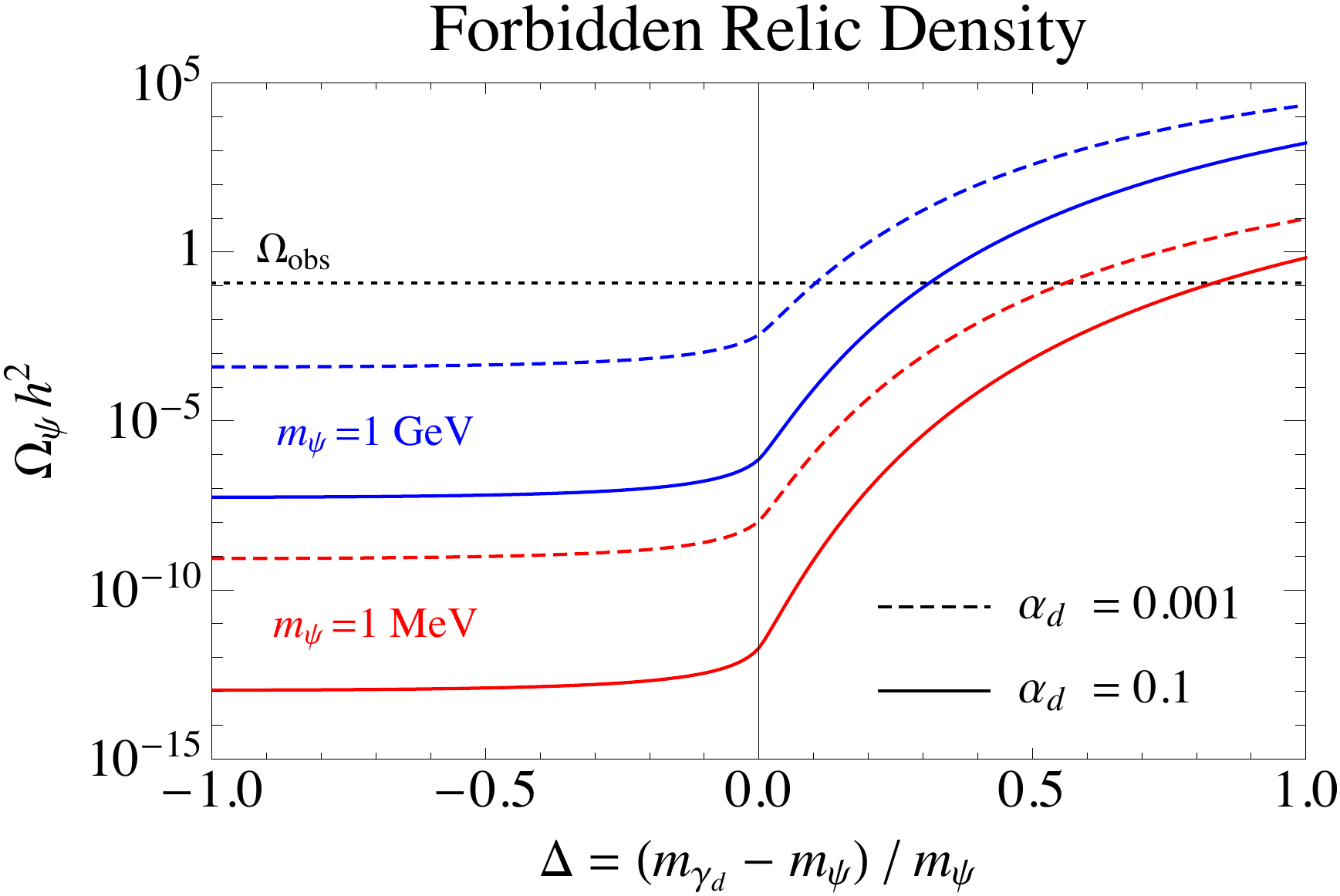}
\end{tabular}
\caption{ \label{fig:FeynOmega}
The {\it left} panel contains Feynman diagrams relevant for (1) the relic density, (2) self-interactions, (3) indirect detection, and (4) direct detection.  The {\it right} panel shows the relic density, $\Omega_\psi h^2$, as a function of the mass splitting $\Delta \equiv (m_{\gamma_d} - m_\psi) / m_\psi$.  The red (blue) curves correspond to $m_\psi = 1$~GeV (MeV) and the solid (dashed) curves correspond to $\alpha_d = 0.1 ~(10^{-3})$.}
\end{figure*}

In this letter we propose a novel framework with naturally light DM\@.   We assume that DM dominantly annihilates into heavier particles: $\psi \bar \psi \rightarrow x y$ with $m_x + m_y > 2 m_\psi$.  Such annihilations are called {\it forbidden channels} because they vanish at zero temperature.  Forbidden channels can proceed at finite temperature in the early universe, due to the thermal tail with high velocity $\psi$'s.  The thermally averaged cross section for forbidden channels is exponentially suppressed, as we derive below.  This exponential suppression allows for a weak scale cross section when $\alpha_d^2 / m_\psi^2$ is much larger than the weak scale, permitting light $m_\psi$ with a sizable $\alpha_d$.  While it is an old observation that forbidden channels can dominate DM annihilations~\cite{Griest:1990kh}, previous work has focused on weak scale DM (see for example Refs.~\cite{Jackson:2009kg,Tulin:2012uq,Jackson:2013pjq,Jackson:2013tca}) and metastable relics~\cite{Pospelov:2010cw}. Forbidden channels remain unexplored as a mechanism for light DM\@.

We call {\it Forbidden Dark Matter} the class of models where the DM relic abundance is dominantly set by forbidden channels. Models of light DM are highly constrained by the Cosmic Microwave Background (CMB), which is sensitive to energy injection into the photon plasma at the recombination epoch~\cite{Padmanabhan:2005es,  Planck:2015xua}.  Forbidden DM evades these constraints because annihilations shut off at low temperatures, $T_{rec} \ll m_x + m_y - 2m_\psi $.  This framework naturally includes large self-interactions, $\psi \psi \rightarrow \psi \psi$, which are unsuppressed and therefore have an exponentially larger cross section than the forbidden annihilation rate.  Large DM self-interactions may help address problems with structure formation in non-interacting DM scenarios~\cite{Spergel:1999mh,deBlok:2009sp,BoylanKolchin:2011de,Rocha:2012jg,Peter:2012jh,Zavala:2012us} and may be indicated by recent cluster observations~\cite{Massey:2015dkw, Kahlhoefer:2015vua}.

For the remainder of this letter, we consider an example model that illustrates the key features of the Forbidden DM framework.  We take the DM to be a Dirac fermion, $\psi$, that is neutral under Standard Model quantum numbers and charged under a hidden $U(1)_d$ gauge symmetry that is broken at low energies.  DM annihilates into hidden photons, $\psi \bar \psi \rightarrow  \gamma_d \gamma_d$ (see the first diagram in Fig.~\ref{fig:FeynOmega}).  The parameters of the model are $(m_\psi, m_{\gamma_d}, \alpha_d)$.  We consider the forbidden portion of parameter space, $m_\psi < m_{\gamma_d}$.  The non-forbidden parameter space of this model was studied by Refs.~\cite{Pospelov:2007mp, Izaguirre:2015yja}.  

The organization of the rest of this letter is as follows.  First, we discuss the computation of the relic density.  Second, we derive the DM self-interaction cross section.  Third, we consider the possibility that the hidden photon couples to the Standard Model (SM) through gauge kinetic mixing, leading to signals in direct and indirect detection.  Finally, we consider the possibility that the hidden sector is thermally decoupled from the SM, allowing for sub-MeV DM masses.  We finish with our conclusions.  In an upcoming paper, we will consider light DM from co-annihilations, where an exponential factor also allows for very light DM~\cite{FutureLightCo}.

{\bf Relic Density:}  The relic density of Forbidden DM is determined by the solution of its Boltzmann equation,
\begin{equation} \label{eq:Boltz}
\dot n_\psi + 3 H n_\psi = - \left< \sigma_{\psi \bar \psi} \, v \right> n_\psi^2 +  \left< \sigma_{\gamma_d \gamma_d} \, v \right> (n_{\gamma_d}^{eq})^2,
\end{equation}
where $n_{\psi, \gamma_d}$ are the number densities, $H$ is the Hubble parameter, $\langle \sigma_{\psi \bar \psi (\gamma_d \gamma_d)} \, v \rangle$ denotes the thermally averaged (inverse-)annihilations, and we have assumed that $\gamma_d$ remains in equilibrium during freeze-out.  The solution is approximately given by Eq.~(\ref{eq:Omega}), with the annihilation rate given by $\left< \sigma_{\psi \bar \psi} \, v \right>$. For simplicity, Eq.~(\ref{eq:Omega}) neglects the dependence on the number of relativistic degrees of freedom and the freeze-out temperature.  These effects are included in our numerical results (for more precise analytic treatments see Refs.~\cite{Kolb:1990vq, Gondolo:1990dk,Steigman:2012nb}).
 
We now introduce a simple prescription for computing the thermal average of the forbidden annihilation rate.
Detailed balance states that the right-hand side of Eq.~(\ref{eq:Boltz}) vanishes in equilibrium, $n_{\psi} = n_{\psi}^{eq}$.  Therefore, the forbidden annihilation rate is related to the rate of the inverse process, which proceeds at 0 temperature, $\left< \sigma_{\gamma_d \gamma_d} \, v \right> \sim \alpha_d^2 / m_{\gamma_d}^2$.  We find,
\begin{equation} \label{eq:ForbSigma}
\left< \sigma_{\psi \bar \psi} \, v \right> = \frac{(n_{\gamma_d}^{eq})^2}{(n_\psi^{eq})^2} \left< \sigma_{\gamma_d \gamma_d} \, v \right>  \approx 8 \pi  f_\Delta \frac{\alpha_d^2}{m_\psi^2} \, e^{- 2 \Delta x},
\end{equation}
where $\Delta \equiv (m_{\gamma_d} - m_\psi) / m_\psi$ is the relative mass splitting, $x \equiv m_\psi / T$, and $f_\Delta \equiv (\Delta^{3/2} (2+\Delta)^{3/2}(2+\Delta (2+\Delta))) / (1+\Delta)^4$. The exponential suppression comes from the form of the equilibrium number density for non-relativistic species, $n_{eq} = g (m T / 2 \pi)^{3/2} \exp(-m/T)$, where $g = 2 \, (3)$ for $\psi \, (\gamma_d)$, and we have assumed zero chemical potential.  Note that the approximation of the forbidden cross section in Ref.~\cite{Griest:1990kh} has an incorrect exponential dependence on $\Delta x$.

\begin{figure*}[tb]
\centering
\includegraphics[width=\linewidth]{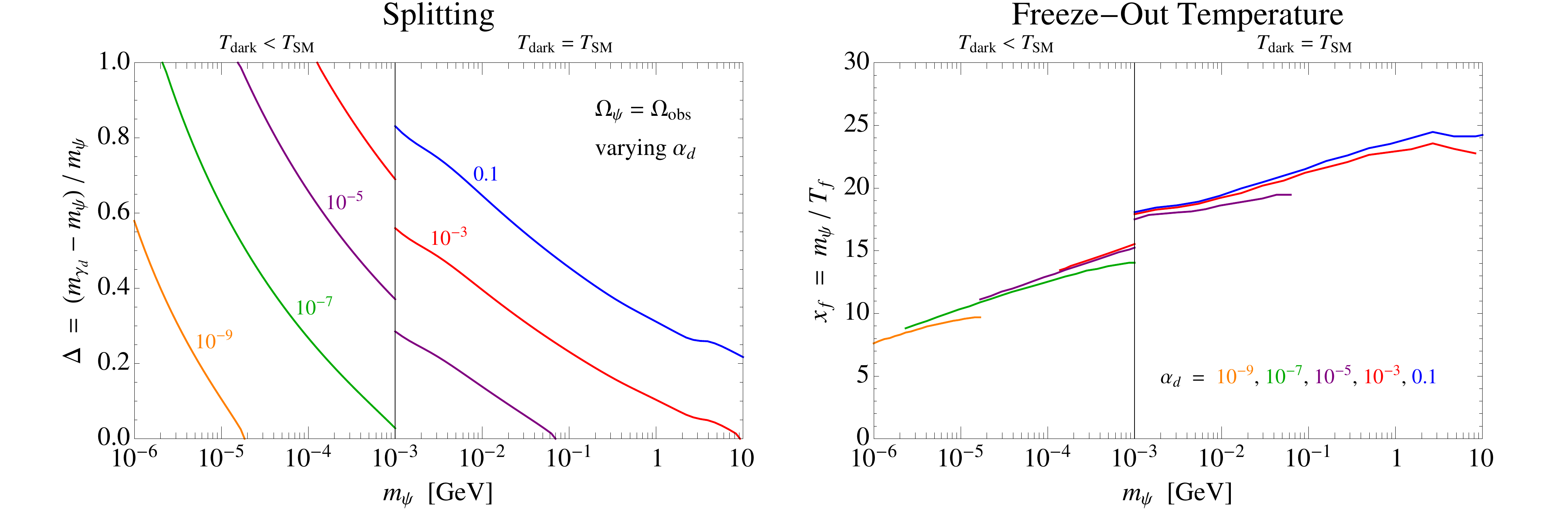}
\caption{ \label{fig:Splitting}
The {\it left} side shows the splitting $\Delta$, versus $m_\psi$, with $\Omega_\psi h^2$ fixed to the observed abundance.  Different curves correspond to different values of $\alpha_d$.  For $m_\psi > 1$~MeV the dark sector temperature equals the SM temperature, while for $m_\psi < 1$~MeV the dark sector is cooler than the SM as described in the Thermally Decoupled Dark Sector section.  The  {\it right} side shows the freeze-out temperature, $x_f \equiv m_\psi / T_f$, as a function of $m_\psi$, with $\Delta$ fixed to the value shown to the left.}
\end{figure*}

We obtain the forbidden relic density by plugging Eq.~(\ref{eq:ForbSigma}) into Eq.~(\ref{eq:Omega}) and integrating the cross section from freeze-out to the present in order to account for annihilations after freeze-out (see for example Ref.~\cite{Gondolo:1990dk}), 
\begin{equation} \label{eq:ForbOmega}
\Omega_\psi h^2 \, \approx \,  0.1 \, g_\Delta(x_f) \, \frac{m_\psi^2/\alpha_d^2}{(20~\mathrm{TeV})^2} \, e^{2 \Delta x_f}  \, ,
\end{equation}
where $x_f \equiv m_\psi / T_f \sim 10-25$ and $g_{\Delta} (x_f) \equiv (4 \pi f_\Delta)^{-1} (1-2 \Delta x_f e^{2 \Delta x_f} \int_{2 \Delta x_f}^\infty t^{-1} e^{-t} dt)^{-1}$ is an $\mathcal{O}(1)$ function. Note that we indicate with $\Omega_\psi h^2$ the total relic density of $\psi$ and $\bar \psi$. Eq.~(\ref{eq:ForbOmega}) shows that the forbidden relic density is exponentially enhanced as $\Delta$ increases.  Equivalently, fixing the relic density to the observed value, the DM mass is exponentially lighter than the weak scale.  

We show the relic density, as a function of $\Delta$, in the right panel of Fig.~\ref{fig:FeynOmega}.  Our numerical results here, and throughout this letter, utilize {\tt MicrOMEGASv4}~\cite{Belanger:2014vza} to solve the Boltzmann equations and we have verified that they agree with Eq.~(\ref{eq:ForbOmega}).  The left of the figure, $\Delta < 0$, corresponds to the conventional case where the relic density is too small for light DM masses.  As we enter the forbidden region, $\Delta > 0$, the relic density exponentially increases until it achieves the correct value.  The standard lore is that forbidden channels are only relevant in highly degenerate scenarios, $\Delta \ll 1$ (this was stated by Ref.~\cite{Griest:1990kh} which implicitly assumes weak scale DM).  However, we see from Fig.~\ref{fig:FeynOmega} that light DM calls for an $\mathcal{O}(1)$ splitting.

On the left side of Fig.~\ref{fig:Splitting}, we show the value of $\Delta$ that corresponds to the observed DM abundance, as a function of the DM mass.  For $m_\psi > 1$~MeV, we assume that the dark sector is in thermal contact with the SM, $T_{\rm dark} = T_{SM}$.  Lighter masses require DM to be thermally decoupled and cooler, $T_{\rm dark} < T_{SM}$, due to constraints on the number of relativistic degrees of freedom from Big Bang Nucleosynthesis (BBN)~\cite{Cyburt:2004yc, Cooke:2013cba} and the CMB~\cite{Planck:2015xua}.  For $m_\psi < 1$~MeV, we adopt a decoupled dark sector scenario, consistent with these constraints, that we describe below.  We find that DM masses down to the keV scale are accommodated (DM with a sub-keV mass is too warm, causing problems with structure formation~\cite{Viel:2005qj, Loeb:2005pm, Hooper:2007tu}).  The right side of Fig.~\ref{fig:Splitting} shows the value of the freeze-out temperature, $x_f \sim 10-25$.

{\bf Self-Interactions:}  Sizable DM self-interactions may leave observable imprints on astrophysical observations~\cite{Spergel:1999mh}.  The self-interaction cross section is constrained by cluster mergers~\cite{Markevitch:2003at,Randall:2007ph,Harvey:2015hha, Clowe:2003tk} and halo shapes~\cite{Rocha:2012jg, Peter:2012jh}, $\sigma_{SI} / m_\psi \lesssim 1~\mathrm{cm}^2/\mathrm{g} \sim 5 \times 10^{-6}~\mathrm{MeV}^{-3}$.  A cross section near this limit, $\sigma_{SI} / m_\psi \gtrsim 0.1~\mathrm{cm}^2/\mathrm{g}$, may help resolve the core-cusp~\cite{deBlok:2009sp} and too big to fail~\cite{BoylanKolchin:2011de} problems of structure formation (see for example Refs.~\cite{Rocha:2012jg,Peter:2012jh,Zavala:2012us}), although baryonic feedback may be important for addressing these problems~\cite{Governato:2009bg,Pontzen:2011ty,Governato:2012fa,Zolotov:2012xd}.  Self-interactions may also be indicated by recent cluster observations~\cite{Massey:2015dkw, Kahlhoefer:2015vua}.  However, an observable self-interacting cross section is orders of magnitude larger than the annihilation cross section at freeze-out, making self-interactions irrelevant for many models of weak scale DM (models with light mediators can be an exception~\cite{Feng:2009hw,Buckley:2009in,Loeb:2010gj}).

\begin{figure*}[tb]
\centering
\includegraphics[width=\linewidth]{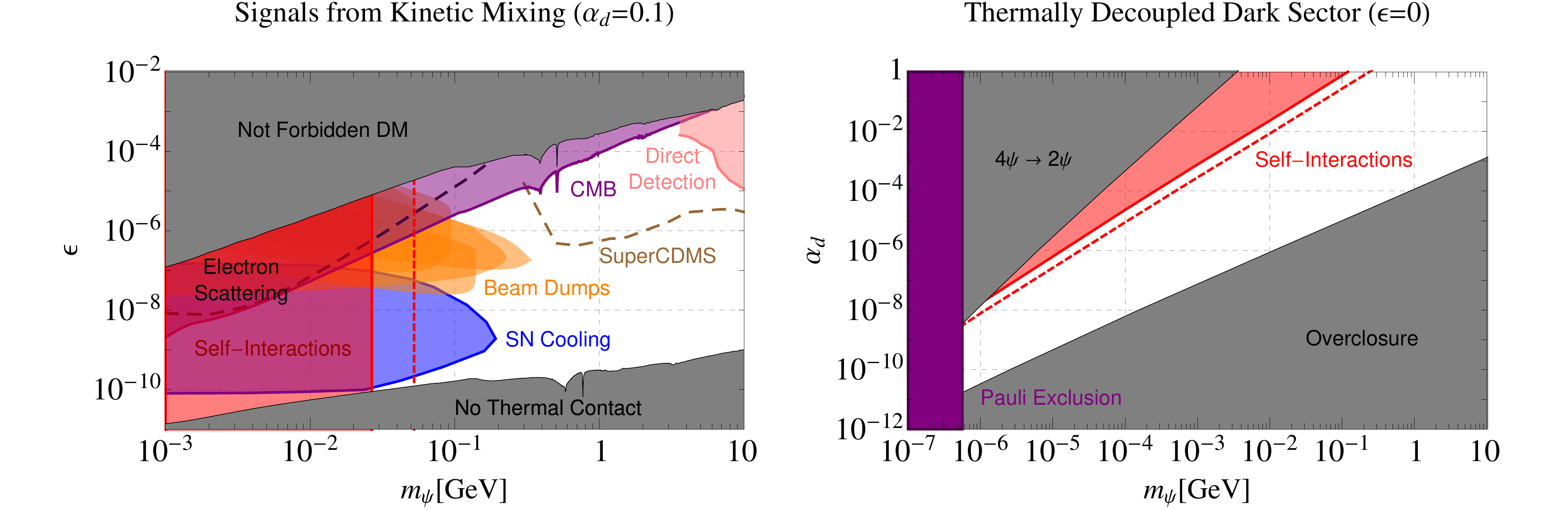}
\caption{\label{fig:pheno}
The {\it left} side shows constraints on the kinetic mixing parameter, $\epsilon$, versus $m_\psi$, with $\Delta$ chosen to match the observed relic density and $\alpha_d = 0.1$.  Annihilation to SM states dominates over forbidden channels in the upper gray region, and the dark sector is thermally decoupled from the SM in the lower gray region.  
Limits are shown from beam dump experiments (orange)~\cite{Bjorken:2009mm, Essig:2010gu, Blumlein:2011mv,Gninenko:2012eq,Andreas:2012mt}, supernovae cooling (blue)~\cite{Bjorken:2009mm,Dent:2012mx}, Planck (purple)~\cite{Planck:2015xua, Slatyer:2015jla}, and direct detection (pink)~\cite{Agnese:2013jaa,Akerib:2013tjd,Agnese:2014aze}.
The dashed brown (black) curves show the projected reach of SuperCDMS SNOLAB~\cite{Cushman:2013zza} (electron scattering~\cite{Essig:2011nj}). The red shaded area (dashed curve) shows the approximate sensitivity of current observations to DM self-interactions, $\sigma_{SI} / m_\psi \gtrsim 1~(0.1)~\mathrm{cm}^2/\mathrm{g}$~\cite{Rocha:2012jg, Peter:2012jh,Markevitch:2003at,Clowe:2003tk,Randall:2007ph,Harvey:2015hha}.  The {\it right} side shows constraints on the dark sector when it is thermally decoupled from the SM\@.  In the upper gray area $4 \psi \rightarrow 2 \psi$ dominates over forbidden annihilations.  In the lower gray area DM is overabundant. The red shading and dashed curve represent the same values for the self-interaction cross section as in the left panel. In the purple shaded area the DM mass is too small to be simultaneously consistent with the Pauli exclusion principle and the densities observed in dwarf spheroidal galaxies~\cite{Tremaine:1979we,Boyarsky:2008ju, Horiuchi:2013noa}.}
\end{figure*}

Forbidden DM naturally achieves a large hierarchy between the self-interaction cross section, which is unsuppressed, and the annihilation rate, which  is exponentially suppressed after thermal averaging.  Self-interactions are velocity-independent and are generated by the second diagram of Fig.~\ref{fig:FeynOmega}, leading to the cross section,
\begin{equation} \label{eq:SI}
\frac{\sigma_{SI}}{m_\psi} = 3\pi h_\Delta \frac{\alpha_d^2}{m_\psi^3} \approx 10~\frac{\mathrm{cm}^2}{\mathrm{g}}  \times \left( \frac{10~\mathrm{MeV}}{m_\psi}\right)^3  \times \left(\frac{\alpha_d}{0.1} \right)^2
\end{equation}
where $h_\Delta \equiv (5+\Delta(2+\Delta)(5 \Delta(2+\Delta)-6)) / \left( (-1+\Delta)^2 (1+\Delta)^4(3+\Delta)^2 \right)$.  For $\alpha_d\approx 0.1$, observable self-interactions are realizable for Forbidden DM lighter than $\sim 100$~MeV, as shown  in the left panel of Fig.~\ref{fig:pheno}. In the right panel we show the relevant parameter space in the $(m_\psi, \alpha_d)$ plane for a dark sector decoupled from the SM\@.

{\bf Signals from Kinetic Mixing:} The sizes of direct and indirect detection signals depend on how the dark sector couples to the SM\@.  We consider the possibility that the dark photon kinetically mixes with the SM photon~\cite{Holdom:1985ag},
\begin{equation} \label{eq:KM}
\mathcal{L} \supset - \frac{\epsilon}{2} \, F_{\mu \nu}^d F^{\mu \nu},
\end{equation}
where any size for $\epsilon$, which characterizes the strength of mixing, is technically natural.  The kinetic mixing is removed (at leading order in $\epsilon$) by shifting the photon, $A^\mu \rightarrow A^\mu - \epsilon \gamma_d^\mu$, inducing a coupling between the dark photon and the electromagnetic current, $\epsilon \gamma_d^\mu J_\mu^{EM}$.  

The left panel of Fig.~\ref{fig:pheno} shows the $(\epsilon, m_\psi)$ plane, fixing $\alpha_d = 0.1$ and choosing the splitting $\Delta$ at each point so that the relic density matches the observed value, as in Fig.~\ref{fig:Splitting}.  In the lower gray region, the dark sector is thermally decoupled from the SM before freeze-out and the cosmology depends on the initial conditions.  We consider this possibility below.   Limits on the dark photon are shown from beam dump experiments~\cite{Bjorken:2009mm, Essig:2010gu, Blumlein:2011mv,Gninenko:2012eq,Andreas:2012mt} and SN1987A cooling~\cite{Bjorken:2009mm,Dent:2012mx}.

Kinetic mixing allows DM to annihilate to charged states, such as electrons, through off-shell dark photons as in the third diagram of Fig.~\ref{fig:FeynOmega}.  The cross section of annihilations into electrons is
\begin{equation} \label{eq:SigmaEps}
\left< \sigma v \right>_\epsilon = \epsilon^2 \frac{16 \pi  \alpha \alpha_d}{m_\psi^2 (3-2\Delta-\Delta^2)^2}.
\end{equation}
This cross section dominates over forbidden annihilations in the upper gray region of Fig.~\ref{fig:pheno}.  By the recombination epoch, forbidden annihilations shut off, but annihilations into charged states are still active and are constrained by CMB observations~\cite{Padmanabhan:2005es,Slatyer:2009yq,Galli:2013dna,Madhavacheril:2013cna}.  The purple region of Fig.~\ref{fig:FeynOmega} is excluded by Planck~\cite{Planck:2015xua}, where we have included the efficiency for annihilations to deposit energy into the photon plasma from Ref.~\cite{Slatyer:2015jla}.  The CMB limit supersedes the present reach of diffuse gamma and X-ray observations~\cite{Essig:2013goa}.

Kinetic mixing also allows DM to scatter against nuclei, as in the fourth diagram of Fig.~\ref{fig:FeynOmega}.  The DM-nucleon cross section is~\cite{Pospelov:2007mp}, 
\begin{equation} \label{eq:SigmaDD}
\sigma_{\psi N} \approx \epsilon^2 \, \frac{16 \pi \alpha \alpha_d \mu_{\psi p}^2}  {m_{\gamma_d}^4} \, \frac{Z^2}{A^2},
\end{equation}
where $\mu_{\psi p} = m_\psi m_p / (m_\psi + m_p)$ is the reduced mass of DM and the proton.  In Fig.~\ref{fig:pheno}, we show the strongest present limits from direct detection, which, moving from heavier to lighter DM mass, come from LUX~\cite{Akerib:2013tjd}, SuperCDMS Soudan~\cite{Agnese:2014aze}, and CDMSlite~\cite{Agnese:2013jaa}.  We also show the projected sensitivity of SuperCDMS SNOLAB~\cite{Cushman:2013zza}, which will probe a significant fraction of parameter space.  DM can also scatter against electrons and we show the estimated reach of a future germanium detector~\cite{Essig:2011nj}, although it is superseded in this model by the Planck constraint.

{\bf Thermally Decoupled Dark Sector:}
We now consider the possibility that the dark sector is thermally decoupled from the SM during freeze-out, $\epsilon \rightarrow 0$.  Our treatment of the relic density assumes that the dark photons remain in equilibrium during freeze-out, with zero chemical potential, as happens if the dark photons are thermalized with radiation.  In the $\epsilon \rightarrow 0$ limit, we assume there is dark radiation, $n$, that couples to the hidden photon,
\begin{equation} \label{eq:DarkRad}
\mathcal{L} \supset  q_n g_d \, \bar n \gamma_\mu n \, \gamma_d^\mu
\end{equation}
where $m_n \ll m_\psi$ and $q_n \ll1$ is the charge of $n$ under the dark force.  We assume that $q_n$ is large enough to keep $\gamma_d$ in equilibrium with $n$ but small enough to prevent $\psi \psi \rightarrow n \bar n$ decays from dominating over forbidden annihilations.  A large range of parameters satisfies these conditions, $10^{-10} \lesssim q_n  \lesssim 10^{-4}$.  For $m_n \lesssim 1$~eV, $n$ is a warm, subdominant, component of DM that contributes less than $10\%$ of the DM energy density, satisfying constraints on warm DM~\cite{Viel:2005qj}.

In general, the dark sector has a different temperature than the SM when the two sectors are thermally decoupled.  We assume that the two sectors begin with a common temperature above the weak scale, $T_0 \gtrsim v$.  Then, the relative temperatures of the two sectors is determined by the requirement that they separately conserve entropy~\cite{Feng:2008mu},
\begin{equation} \label{eq:DarkTemp}
\frac{T_{\rm dark}}{T_{SM}} = \left(\frac{g_{*S}^{SM}(T_{SM})}{g_{*S}^{SM}(T_0)}   \frac{g_{*S}^{\rm dark}(T_0)}{g_{*S}^{\rm dark}(T_{\rm dark})} \right)^{1/3}.
\end{equation}
In our model, the hidden sector becomes cooler than the SM because more states freeze-out in the SM sector.  At low temperatures, $T \ll m_\psi$, $T_{\rm dark} \approx 0.5 \, T_{SM}$.  Because of the smaller dark temperature, the hidden sector is consistent with constraints on the number of relativistic degrees of freedom from BBN~\cite{Cyburt:2004yc, Cooke:2013cba} and the CMB~\cite{Planck:2015xua}, including when $m_\psi \ll T_{BBN} \sim 1$~MeV\@. We computed these constraints in the presence of a dark Higgs with the same mass as the dark photon.

To the right of Fig.~\ref{fig:pheno}, we show the parameter space of the decoupled scenario in the $(m_\psi, \alpha_d)$ plane, fixing $\Delta$ to match the observed relic density.  In the upper gray area, $m_{\gamma_d} > 2 m_\psi$ and $4 \psi \rightarrow 2 \psi$ dominates over forbidden annihilations.   In the lower gray region, the abundance is too large for any choice of $\Delta$.  The shaded red region (red dashed line) indicates the limit (approximate observable reach) for self-interactions, $\sigma_{SI} / m_\psi \gtrsim 1\,  (0.1)~\mathrm{cm}^2/\mathrm{g}$. Very light fermionic DM, $m_\psi \lesssim $~keV, is not allowed because the Pauli exclusion principle is inconsistent with observed densities of dwarf spheroidal galaxies (purple shading)~\cite{Tremaine:1979we,Boyarsky:2008ju, Horiuchi:2013noa}. A comparable bound is obtained from Lyman-$\alpha$ observations~\cite{Viel:2005qj, Loeb:2005pm, Hooper:2007tu} and also applies to bosonic DM\@. However it can be relaxed if we give up the assumption of a common temperature for the two sectors above the weak scale.

{\bf Conclusions:}
In this letter we have proposed the Forbidden DM framework.  DM may dominantly annihilate into heavier states, such that the exponential suppression of the thermally averaged cross section allows for DM that is exponentially lighter than the weak scale.  Forbidden channels shut off at low temperatures, naturally evading CMB constraints.  Self-interactions are unsuppressed, and potentially observable for Forbidden DM lighter than about 100 MeV\@.  We have illustrated Forbidden DM with a sample model, where fermionic DM annihilates into dark photons.  We leave consideration of more forbidden models for future work, including scenarios where DM annihilates through $p$-wave processes or is a composite state of a strongly coupled hidden sector.

\begin{acknowledgements}
{\it\bf Acknowledgments:} 
We thank James Bullock, Rouven Essig, Tracy Slatyer, and Neal Weiner for helpful discussions.  JTR thanks Tel Aviv University for hospitality while this work was completed.  RTD was supported by the NSF grant PHY-0907744.
\end{acknowledgements}



\begin{thebibliography}{0}


\bibitem{Kolb:1990vq} 
  E.~W.~Kolb and M.~S.~Turner,
  Front.\ Phys.\  {\bf 69}, 1 (1990).
  
\bibitem{Cushman:2013zza} 
  P.~Cushman, C.~Galbiati, D.~N.~McKinsey, H.~Robertson, T.~M.~P.~Tait, D.~Bauer, A.~Borgland and B.~Cabrera {\it et al.},
  arXiv:1310.8327 [hep-ex].
  
\bibitem{Feng:2008ya} 
  J.~L.~Feng and J.~Kumar,
  Phys.\ Rev.\ Lett.\  {\bf 101}, 231301 (2008)
  [arXiv:0803.4196 [hep-ph]].
  
\bibitem{Nussinov:1985xr} 
  S.~Nussinov,
  Phys.\ Lett.\ B {\bf 165}, 55 (1985).
  
\bibitem{Kaplan:2009ag} 
  D.~E.~Kaplan, M.~A.~Luty and K.~M.~Zurek,
  Phys.\ Rev.\ D {\bf 79}, 115016 (2009)
  [arXiv:0901.4117 [hep-ph]].
  
\bibitem{Carlson:1992fn} 
  E.~D.~Carlson, M.~E.~Machacek and L.~J.~Hall,
  Astrophys.\ J.\  {\bf 398}, 43 (1992).
  
\bibitem{Hochberg:2014dra} 
  Y.~Hochberg, E.~Kuflik, T.~Volansky and J.~G.~Wacker,
  Phys.\ Rev.\ Lett.\  {\bf 113}, 171301 (2014)
  [arXiv:1402.5143 [hep-ph]].


\bibitem{Griest:1990kh} 
  K.~Griest and D.~Seckel,
  Phys.\ Rev.\ D {\bf 43}, 3191 (1991).
  
\bibitem{Jackson:2009kg} 
  C.~B.~Jackson, G.~Servant, G.~Shaughnessy, T.~M.~P.~Tait and M.~Taoso,
  JCAP {\bf 1004}, 004 (2010)
  [arXiv:0912.0004 [hep-ph]].
  
  
  
\bibitem{Tulin:2012uq} 
  S.~Tulin, H.~B.~Yu and K.~M.~Zurek,
  Phys.\ Rev.\ D {\bf 87}, no. 3, 036011 (2013)
  [arXiv:1208.0009 [hep-ph]].
  
\bibitem{Jackson:2013pjq} 
  C.~B.~Jackson, G.~Servant, G.~Shaughnessy, T.~M.~P.~Tait and M.~Taoso,
  JCAP {\bf 1307}, 021 (2013)
  [arXiv:1302.1802 [hep-ph]].
  
\bibitem{Jackson:2013tca} 
  C.~B.~Jackson, G.~Servant, G.~Shaughnessy, T.~M.~P.~Tait and M.~Taoso,
  JCAP {\bf 1307}, 006 (2013)
  [arXiv:1303.4717 [hep-ph]].
  
\bibitem{Pospelov:2010cw} 
  M.~Pospelov and J.~Pradler,
  Phys.\ Rev.\ D {\bf 82}, 103514 (2010)
  [arXiv:1006.4172 [hep-ph]].
  
\bibitem{Padmanabhan:2005es} 
  N.~Padmanabhan and D.~P.~Finkbeiner,
  Phys.\ Rev.\ D {\bf 72}, 023508 (2005)
  [astro-ph/0503486].
 
\bibitem{Planck:2015xua} 
  P.~A.~R.~Ade {\it et al.}  [Planck Collaboration],
  arXiv:1502.01589 [astro-ph.CO].
  
\bibitem{Spergel:1999mh} 
  D.~N.~Spergel and P.~J.~Steinhardt,
  Phys.\ Rev.\ Lett.\  {\bf 84}, 3760 (2000)
  [astro-ph/9909386].
  
\bibitem{deBlok:2009sp} 
  W.~J.~G.~de Blok,
  Adv.\ Astron.\  {\bf 2010}, 789293 (2010)
  [arXiv:0910.3538 [astro-ph.CO]].

\bibitem{BoylanKolchin:2011de} 
  M.~Boylan-Kolchin, J.~S.~Bullock and M.~Kaplinghat,
  Mon.\ Not.\ Roy.\ Astron.\ Soc.\  {\bf 415}, L40 (2011)
  [arXiv:1103.0007 [astro-ph.CO]].

\bibitem{Rocha:2012jg} 
  M.~Rocha, A.~H.~G.~Peter, J.~S.~Bullock, M.~Kaplinghat, S.~Garrison-Kimmel, J.~Onorbe and L.~A.~Moustakas,
  Mon.\ Not.\ Roy.\ Astron.\ Soc.\  {\bf 430}, 81 (2013)
  [arXiv:1208.3025 [astro-ph.CO]].
    
\bibitem{Peter:2012jh} 
  A.~H.~G.~Peter, M.~Rocha, J.~S.~Bullock and M.~Kaplinghat,
  Mon.\ Not.\ Roy.\ Astron.\ Soc.\  {\bf 430}, 105 (2013)
  [arXiv:1208.3026 [astro-ph.CO]].
  
\bibitem{Zavala:2012us} 
  J.~Zavala, M.~Vogelsberger and M.~G.~Walker,
  Monthly Notices of the Royal Astronomical Society: Letters {\bf 431}, L20 (2013)
  [arXiv:1211.6426 [astro-ph.CO]].
  
\bibitem{Massey:2015dkw} 
  R.~Massey, L.~Williams, R.~Smit, M.~Swinbank, T.~D.~Kitching, D.~Harvey, M.~Jauzac and H.~Israel {\it et al.},
  Mon.\ Not.\ Roy.\ Astron.\ Soc.\  {\bf 449}, 3393 (2015)
  [arXiv:1504.03388 [astro-ph.CO]].
  \bibitem{Kahlhoefer:2015vua} 
  F.~Kahlhoefer, K.~Schmidt-Hoberg, J.~Kummer and S.~Sarkar,
  arXiv:1504.06576 [astro-ph.CO].

  

  
    
\bibitem{Pospelov:2007mp} 
  M.~Pospelov, A.~Ritz and M.~B.~Voloshin,
  Phys.\ Lett.\ B {\bf 662}, 53 (2008)
  [arXiv:0711.4866 [hep-ph]].
  
\bibitem{Izaguirre:2015yja} 
  E.~Izaguirre, G.~Krnjaic, P.~Schuster and N.~Toro,
  arXiv:1505.00011 [hep-ph].
  
\bibitem{FutureLightCo} 
  R.~T.~D'Agnolo and J.~T.~Ruderman,
  {\it to appear}.
  
\bibitem{Gondolo:1990dk} 
  P.~Gondolo and G.~Gelmini,
  Nucl.\ Phys.\ B {\bf 360}, 145 (1991).
  
\bibitem{Steigman:2012nb} 
  G.~Steigman, B.~Dasgupta and J.~F.~Beacom,
  Phys.\ Rev.\ D {\bf 86}, 023506 (2012)
  [arXiv:1204.3622 [hep-ph]].
  
\bibitem{Belanger:2014vza} 
  G.~Belanger, F.~Boudjema, A.~Pukhov and A.~Semenov,
  arXiv:1407.6129 [hep-ph].
  

\bibitem{Cyburt:2004yc} 
  R.~H.~Cyburt, B.~D.~Fields, K.~A.~Olive and E.~Skillman,
  Astropart.\ Phys.\  {\bf 23}, 313 (2005)
  [astro-ph/0408033].
  
\bibitem{Cooke:2013cba} 
  R.~Cooke, M.~Pettini, R.~A.~Jorgenson, M.~T.~Murphy and C.~C.~Steidel,
  arXiv:1308.3240 [astro-ph.CO].



\bibitem{Viel:2005qj} 
  M.~Viel, J.~Lesgourgues, M.~G.~Haehnelt, S.~Matarrese and A.~Riotto,
  Phys.\ Rev.\ D {\bf 71}, 063534 (2005)
  [astro-ph/0501562].
  \bibitem{Loeb:2005pm} 
  A.~Loeb and M.~Zaldarriaga,
  Phys.\ Rev.\ D {\bf 71}, 103520 (2005)
  [astro-ph/0504112].
  \bibitem{Hooper:2007tu} 
  D.~Hooper, M.~Kaplinghat, L.~E.~Strigari and K.~M.~Zurek,
  Phys.\ Rev.\ D {\bf 76}, 103515 (2007)
  [arXiv:0704.2558 [astro-ph]].
  
\bibitem{Tremaine:1979we} 
  S.~Tremaine and J.~E.~Gunn,
  Phys.\ Rev.\ Lett.\  {\bf 42}, 407 (1979).
  
\bibitem{Boyarsky:2008ju} 
  A.~Boyarsky, O.~Ruchayskiy and D.~Iakubovskyi,
  JCAP {\bf 0903}, 005 (2009)
  [arXiv:0808.3902 [hep-ph]].
  
\bibitem{Horiuchi:2013noa} 
  S.~Horiuchi, P.~J.~Humphrey, J.~Onorbe, K.~N.~Abazajian, M.~Kaplinghat and S.~Garrison-Kimmel,
  Phys.\ Rev.\ D {\bf 89}, no. 2, 025017 (2014)
  [arXiv:1311.0282 [astro-ph.CO]].
  
\bibitem{Markevitch:2003at} 
  M.~Markevitch, A.~H.~Gonzalez, D.~Clowe, A.~Vikhlinin, L.~David, W.~Forman, C.~Jones and S.~Murray {\it et al.},
  Astrophys.\ J.\  {\bf 606}, 819 (2004)
  [astro-ph/0309303].
  
    \bibitem{Clowe:2003tk} 
  D.~Clowe, A.~Gonzalez and M.~Markevitch,
  Astrophys.\ J.\  {\bf 604}, 596 (2004)
  [astro-ph/0312273].
  
\bibitem{Randall:2007ph} 
  S.~W.~Randall, M.~Markevitch, D.~Clowe, A.~H.~Gonzalez and M.~Bradac,
  Astrophys.\ J.\  {\bf 679}, 1173 (2008)
  [arXiv:0704.0261 [astro-ph]].

\bibitem{Harvey:2015hha} 
  D.~Harvey, R.~Massey, T.~Kitching, A.~Taylor and E.~Tittley,
  Science {\bf 347}, no. 6229, 1462 (2015)
  [arXiv:1503.07675 [astro-ph.CO]].
  
\bibitem{Governato:2009bg} 
  F.~Governato, C.~Brook, L.~Mayer, A.~Brooks, G.~Rhee, J.~Wadsley, P.~Jonsson and B.~Willman {\it et al.},
  Nature {\bf 463}, 203 (2010)
  [arXiv:0911.2237 [astro-ph.CO]].
  
\bibitem{Pontzen:2011ty} 
  A.~Pontzen and F.~Governato,
  Mon.\ Not.\ Roy.\ Astron.\ Soc.\  {\bf 421}, 3464 (2012)
  [arXiv:1106.0499 [astro-ph.CO]].
  
\bibitem{Governato:2012fa} 
  F.~Governato, A.~Zolotov, A.~Pontzen, C.~Christensen, S.~H.~Oh, A.~M.~Brooks, T.~Quinn and S.~Shen {\it et al.},
  Mon.\ Not.\ Roy.\ Astron.\ Soc.\  {\bf 422}, 1231 (2012)
  [arXiv:1202.0554 [astro-ph.CO]].
  
\bibitem{Zolotov:2012xd} 
  A.~Zolotov, A.~M.~Brooks, B.~Willman, F.~Governato, A.~Pontzen, C.~Christensen, A.~Dekel and T.~Quinn {\it et al.},
  Astrophys.\ J.\  {\bf 761}, 71 (2012)
  [arXiv:1207.0007 [astro-ph.CO]].
 
  
\bibitem{Feng:2009hw} 
  J.~L.~Feng, M.~Kaplinghat and H.~B.~Yu,
  Phys.\ Rev.\ Lett.\  {\bf 104}, 151301 (2010)
  [arXiv:0911.0422 [hep-ph]].
  
\bibitem{Buckley:2009in} 
  M.~R.~Buckley and P.~J.~Fox,
  Phys.\ Rev.\ D {\bf 81}, 083522 (2010)
  [arXiv:0911.3898 [hep-ph]].
  
\bibitem{Loeb:2010gj} 
  A.~Loeb and N.~Weiner,
  Phys.\ Rev.\ Lett.\  {\bf 106}, 171302 (2011)
  [arXiv:1011.6374 [astro-ph.CO]].
  
 
  
\bibitem{Holdom:1985ag} 
  B.~Holdom,
  Phys.\ Lett.\ B {\bf 166}, 196 (1986).
  
\bibitem{Bjorken:2009mm} 
  J.~D.~Bjorken, R.~Essig, P.~Schuster and N.~Toro,
  Phys.\ Rev.\ D {\bf 80}, 075018 (2009)
  [arXiv:0906.0580 [hep-ph]].
  
\bibitem{Essig:2010gu} 
  R.~Essig, R.~Harnik, J.~Kaplan and N.~Toro,
  Phys.\ Rev.\ D {\bf 82}, 113008 (2010)
  [arXiv:1008.0636 [hep-ph]].
  
\bibitem{Blumlein:2011mv} 
  J.~Blumlein and J.~Brunner,
  Phys.\ Lett.\ B {\bf 701}, 155 (2011)
  [arXiv:1104.2747 [hep-ex]].
  
\bibitem{Gninenko:2012eq} 
  S.~N.~Gninenko,
  Phys.\ Lett.\ B {\bf 713}, 244 (2012)
  [arXiv:1204.3583 [hep-ph]].
  
\bibitem{Andreas:2012mt} 
  S.~Andreas, C.~Niebuhr and A.~Ringwald,
  Phys.\ Rev.\ D {\bf 86}, 095019 (2012)
  [arXiv:1209.6083 [hep-ph]].
  
\bibitem{Dent:2012mx} 
  J.~B.~Dent, F.~Ferrer and L.~M.~Krauss,
  arXiv:1201.2683 [astro-ph.CO].
  
\bibitem{Slatyer:2015jla} 
  T.~R.~Slatyer,
  arXiv:1506.03811 [hep-ph].
  
\bibitem{Agnese:2013jaa} 
  R.~Agnese {\it et al.}  [SuperCDMS Collaboration],
  Phys.\ Rev.\ Lett.\  {\bf 112}, no. 4, 041302 (2014)
  [arXiv:1309.3259 [physics.ins-det]].
  
\bibitem{Akerib:2013tjd} 
  D.~S.~Akerib {\it et al.}  [LUX Collaboration],
  Phys.\ Rev.\ Lett.\  {\bf 112}, 091303 (2014)
  [arXiv:1310.8214 [astro-ph.CO]].
  
\bibitem{Agnese:2014aze} 
  R.~Agnese {\it et al.}  [SuperCDMS Collaboration],
  Phys.\ Rev.\ Lett.\  {\bf 112}, no. 24, 241302 (2014)
  [arXiv:1402.7137 [hep-ex]].
  
\bibitem{Essig:2011nj} 
  R.~Essig, J.~Mardon and T.~Volansky,
  Phys.\ Rev.\ D {\bf 85}, 076007 (2012)
  [arXiv:1108.5383 [hep-ph]].
  
\bibitem{Slatyer:2009yq} 
  T.~R.~Slatyer, N.~Padmanabhan and D.~P.~Finkbeiner,
  Phys.\ Rev.\ D {\bf 80}, 043526 (2009)
  [arXiv:0906.1197 [astro-ph.CO]].
  
\bibitem{Galli:2013dna} 
  S.~Galli, T.~R.~Slatyer, M.~Valdes and F.~Iocco,
  Phys.\ Rev.\ D {\bf 88}, 063502 (2013)
  [arXiv:1306.0563 [astro-ph.CO]].
  
\bibitem{Madhavacheril:2013cna} 
  M.~S.~Madhavacheril, N.~Sehgal and T.~R.~Slatyer,
  Phys.\ Rev.\ D {\bf 89}, 103508 (2014)
  [arXiv:1310.3815 [astro-ph.CO]].
  

  
\bibitem{Essig:2013goa} 
  R.~Essig, E.~Kuflik, S.~D.~McDermott, T.~Volansky and K.~M.~Zurek,
  JHEP {\bf 1311}, 193 (2013)
  [arXiv:1309.4091 [hep-ph]].

  
\bibitem{Feng:2008mu} 
  J.~L.~Feng, H.~Tu and H.~B.~Yu,
  JCAP {\bf 0810}, 043 (2008)
  [arXiv:0808.2318 [hep-ph]].
  
 
 
 \end{thebibliography}
\end{document}